\documentclass[lettersize,journal]{IEEEtran}
\usepackage{amsmath,amsfonts}
\usepackage{algorithmic}
\usepackage{algorithm}
\usepackage{array}
\usepackage[caption=false,font=normalsize,labelfont=sf,textfont=sf]{subfig}
\usepackage{textcomp}
\usepackage{stfloats}
\usepackage{url}
\usepackage{verbatim}
\usepackage{graphicx}
\usepackage{cite}
\usepackage[T1]{fontenc}
\usepackage{ragged2e}

\begin{document}

\title{A Wearable Gait Monitoring System for 17 Gait Parameters Based on Computer Vision}

\author{Jiangang Chen,~\IEEEmembership{Graduate Student Member,~IEEE,} Yung-Hong Sun, Kristen Pickett, Barbara King, Yu Hen Hu,~\IEEEmembership{Life Fellow,~IEEE,} Hongrui Jiang,~\IEEEmembership{Fellow,~IEEE}
\thanks{This Manuscript received on 17 November 2024. The computer vision algorithm utilized in this work was separately developed under a grant by the National Institute of Biomedical Imaging and Bioengineering of the U.S. National Institutes of Health (Grant number R01EB019460).}
\thanks{Jiangang Chen, Yung-Hong Sun, Yu Hen Hu, and Hongrui Jiang are with the Department of Electrical and Computer Engineering, University of Wisconsin–Madison, Madison, WI 53706 USA (email: {jiangang.chen; ysun376; yhhu; hongruijiang}@wisc.edu).}
\thanks{Kristen Pickett is with Department of Kinesiology, University of Wisconsin–Madison, Madison, WI 53706 USA (email: kristen.pickett@wisc.edu).}
\thanks{Barbara King is with School of Nursing, University of Wisconsin–Madison, Madison, WI 53706 USA (email: bjking2@wisc.edu).}}



\maketitle

\begin{abstract}
We developed a shoe-mounted gait monitoring system capable of tracking up to 17 gait parameters, including gait length, step time, stride velocity, and others. The system employs a stereo camera mounted on one shoe to track a marker placed on the opposite shoe, enabling the estimation of spatial gait parameters. Additionally, a Force Sensitive Resistor (FSR) affixed to the heel of the shoe, combined with a custom-designed algorithm, is utilized to measure temporal gait parameters. Through testing on multiple participants and comparison with the gait mat, the proposed gait monitoring system exhibited notable performance, with the accuracy of all measured gait parameters exceeding 93.61\%. The system also demonstrated a low drift of 4.89\% during long-distance walking. A gait identification task conducted on participants using a trained Transformer model achieved 95.7\% accuracy on the dataset collected by the proposed system, demonstrating that our hardware has the potential to collect long-sequence gait data suitable for integration with current Large Language Models (LLMs). The system is cost-effective, user-friendly, and well-suited for real-life measurements.
\end{abstract}

\begin{IEEEkeywords}
Gait Monitoring; Wearable Device; Computer Vision; Stereo Camera; Transformer Model; Gait Identification.
\end{IEEEkeywords}

\section{Introduction}
\IEEEPARstart{T}{he} gait is a critical predictor of physical functional decline \cite{ieee1}. Gait disturbances are among the most common and disabling symptoms of various diseases, including Parkinson's Disease (PD) \cite{ieee2} and Multiple Sclerosis (MS) \cite{ieee3}. Bradykinesia is the predominant symptom in early PD \cite{ieee4}, characterized by reduced arm swing during walking, diminished walking speed, shortened stride length, increased step frequency, and disturbances in double limb support and gait rhythm \cite{ieee5}. MS is a degenerative, autoimmune disease affecting the central nervous system, often resulting in a persistent decline in walking ability \cite{ieee6}. Consequently, gait parameters, such as rhythm, step length, and step time, are typically reduced in comparison to those observed in healthy individuals \cite{ieee7}. These conditions are prevalent among older adults, increasing their risk of falls, fractures, and disability \cite{ieee8}, resulting in a lower quality of life, limited ability to conduct daily activities, increased morbidity, and higher living expenses. Continuous and regular monitoring of activity levels and gait parameters in elderly individuals could support the diagnosis and assessment of gait disturbances, providing valuable insights into the severity of underlying diseases \cite{ieee9}.

\begin{figure}[!t]
\centering
\includegraphics[width=3.47in, height=1.24in]{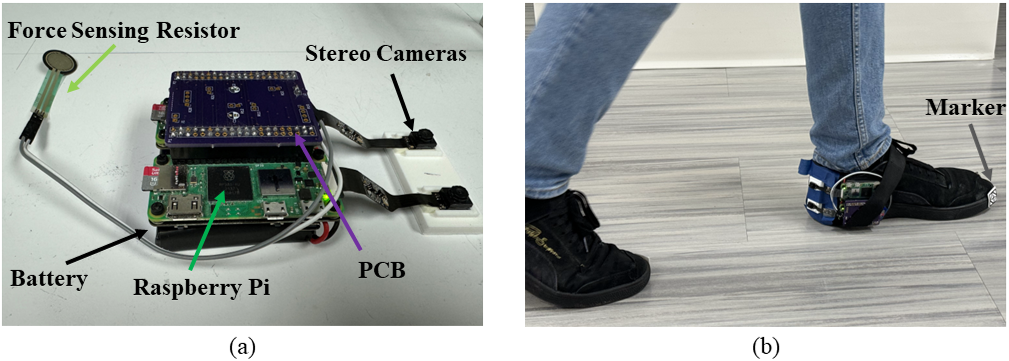}
\caption{(a) Image of the gait monitoring system with all the components, including Raspberry Pis, batteries, FSR, stereo cameras, and PCB; (b) A volunteer walking while wearing the system attached to the side of the shoe, with a custom-designed marker placed on the tip.}
\label{fig_1}
\end{figure}
In clinical practice, commonly used methods for assessing gait performance include visual observation, rating scales, and timed or distance-based tests \cite{ieee10}. The accuracy of subjective results obtained through visual observation relies heavily on the experience of the clinician conducting the gait assessment \cite{ieee11}. In the early stages of PD, visual gait assessments may yield uncertain results, primarily because observed characteristics such as slow gait and short stride length may also be associated with aging, depressive mood, or other conditions \cite{ieee12}. The use of rating scales introduces clinical discrepancies \cite{ieee13}, as each component of the scale represents an ordinal measure rather than a precise interval change \cite{ieee14}. Similarly, outcome measures from timed or distance-based tests lack sensitivity to subtle changes in gait disturbances over the course of trials.

With advancements in technology, several methods have emerged for monitoring gait parameters to assist experts in diagnosis. Current commercial technologies in clinical practice can be categorized into three areas: wearable devices, pressure measurement mats, and motion capture systems \cite{ieee15}. The inertial measurement unit (IMU) \cite{ieee16}, a widely used wearable device, can be placed at the joints, such as the ankle, knee, and hip, to acquire spatiotemporal gait parameters. However, it requires the simultaneous use of multiple IMUs (up to 8 sensors \cite{ieee17}) on joints, reducing user-friendliness. Moreover, due to the challenges of directly correlating IMU signals with known gait characteristics, manual analysis of numerous observed IMU signals is often necessary \cite{ieee18}. Other problems include long-term drift, magnetic interference, and inconsistency \cite{ieee19}. Similarly, floor sensing products and smart insoles integrate pressure measurement technologies, including capacitive, resistive, piezoelectric, and piezoresistive sensors. Floor sensing products, such as gait pressure measurement mats \cite{ieee20}, priced at around 25,000 USD \cite{ieee21}, are cumbersome and limited to capturing only a few steps within a restricted space, resulting in missed gait patterns. Smart insoles face challenges in long-term reliability due to the limited lifespan of sensors and the absence of spatial parameter measurements \cite{ieee22}\cite{ieee23}. Traditional motion capture systems rely on multiple camera system setups and the placement of markers to estimate the positions of anatomical segments. The accuracy of these systems is constrained by the simplistic foot model, as optoelectronic measurement systems can only differentiate a limited number of markers \cite{ieee24}. Motion capture systems that use convolutional neural networks require powerful computational resources \cite{ieee25}. Manual intervention is needed to fix mismatched feature points \cite{ieee26} to describe the human skeleton, and participants have to walk perpendicularly at a fixed depth relative to the camera for spatial parameter measurements \cite{ieee27}.

Generally, existing methods fail to achieve the following simultaneously: (a) accurately measure a wide range of spatiotemporal gait properties; (b) offer low-drift, highly reproducible measurements suitable for long-term, long-distance use; and (c) remain user-friendly, cost-effective, and suitable for real-world applications.

We propose a novel shoe-mounted gait monitoring system based on the stereo camera, capable of measuring up to 17 gait parameters with high accuracy and demonstrate a prototype. The measured gait parameters are highly reproducible and exhibit low signal drift. The system accommodates walks longer than typical gait mats, enabling the collection of a richer gait dataset. This dataset can be integrated into current LLMs to support the diagnosis of gait-related disorders. As shown in Fig. 1, by integrating self-designed markers with a trained object detection model, the large field-of-view (FoV) cameras can capture markers across a wide scene within 0.1 seconds. Subsequently, our system utilizes on-body computer vision processing to detect markers in time-stamped stereo images, enabling the calculation of gait parameters. The system is low-cost (less than \$200), easy to use, and suitable for real-life measurements in both hospital and home settings, all while providing up to 17 gait parameter measurements with competitive accuracy compared to state-of-the-art technologies. Our contributions are as follows:

\begin{itemize}
\item{Development and demonstration of a new gait monitoring system that can measure up to 17 gait parameters with high accuracy.}
\item{The system exhibits low drift for long-distance walks.}
\item{The system’s measured long sequences of gait data can be integrated with LLMs to identify individuals.}
\end{itemize}

This paper is organized as follows. Section II provides a detailed description of the hardware implementation and the calibration process of the stereo cameras. Section III outlines the definitions of 17 gait parameters and the algorithms used to calculate temporal gait parameters. Section IV illustrates the image processing, the stereo algorithm for spatial gait parameters, pseudocode, and personal identification. Finally, the experiment descriptions, results, and discussions are presented in Sections V, VI, and VII, respectively.

\section{Hardware}
This section provides a comprehensive description of the hardware details of the designed gait monitoring system, the procedure of stereo camera calibration, the camera settings, and the approach for installing the device on participants.
\begin{figure}[H]
\centering
\includegraphics[width=3.44in, height=2.11in]{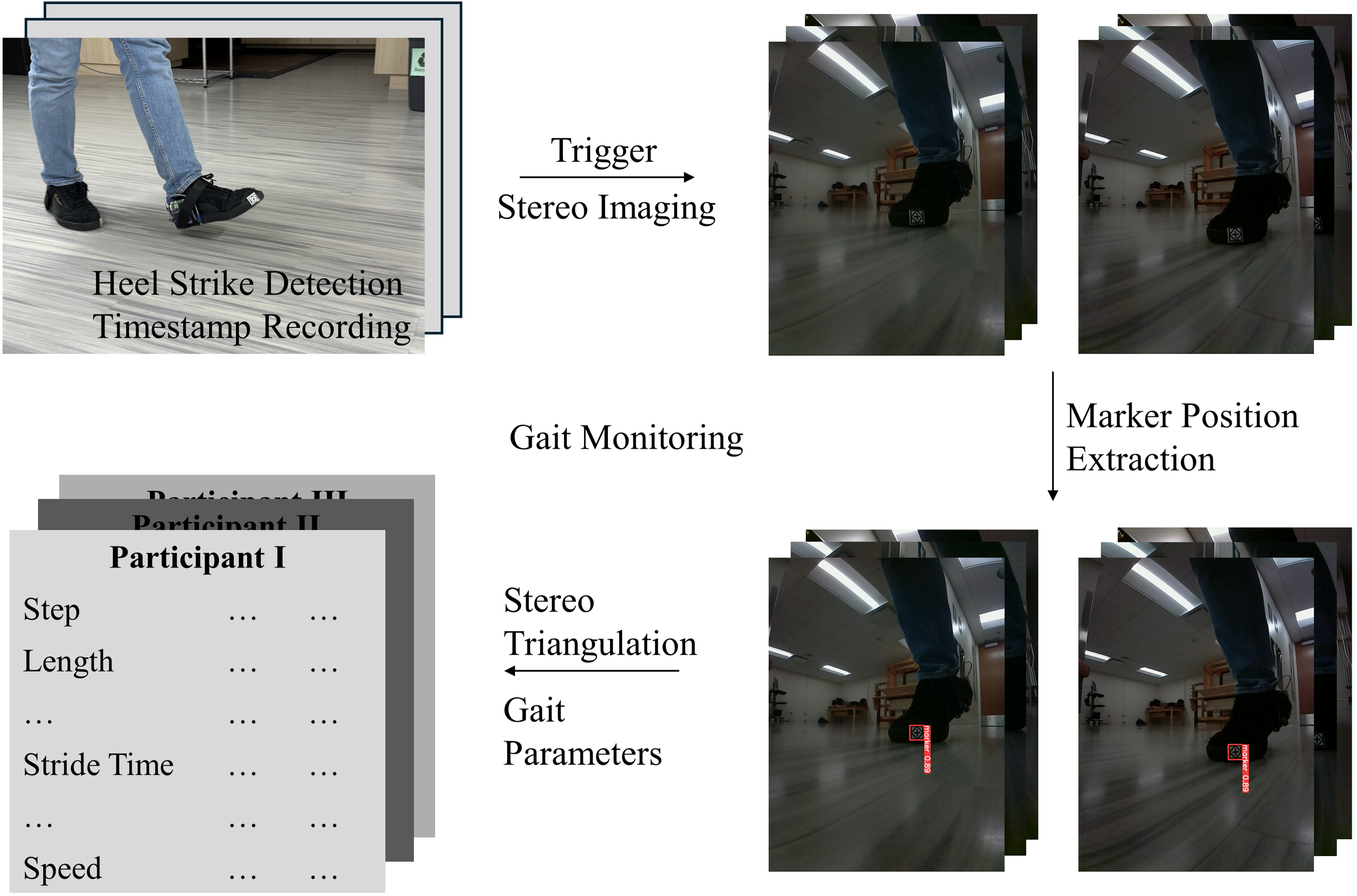}
\caption{Diagram illustrating the overall operation of the gait monitoring system. The large FoV cameras capture a stereo image and record a timestamp each time a heel strike occurs. Subsequently, marker position extraction and stereo triangulation, along with the timestamp, are used to calculate gait parameters.}
\label{fig_2}
\end{figure}
\subsection{Hardware Implementation}
The system includes two wearables, one for each foot. Each wearable is equipped with a pair of Raspberry Pi Zero 2 W microcontrollers, large FoV IR cameras, 1200mAh portable batteries, a polymer force sensing resistor (FSR 402 from Interlink Electronics), and a self-designed marker to form the gait monitoring system. As demonstrated in Fig. 2, the 120° FoV cameras are triggered by an FSR placed at the heel of the foot. During the heel strike phases, the stereo cameras capture images of the marker adhered to the tip of the other shoe and record the timestamp. The extracted marker pixel positions, along with the timestamp, are used for triangulation and calculation of gait parameters.

The two Raspberry Pis in each wearable are physically and electrically connected via a PCB. Two female pin header socket connector strips on either side of the board allow the Raspberry Pis to be securely inserted and easily removed, maintaining the compact size of the wearable. Furthermore, the PCB design focuses on detecting gait events and facilitating communication between the Raspberry Pis, thereby ensuring synchronized image capture by the stereo cameras. The designed force sensing circuit operates on the same principle as a low-pass filter, where the FSR functions as a variable resistive element and a 100nF multilayer ceramic capacitor serves as the output load. The time taken for the capacitor to reach the high state indicates the approximate magnitude of the FSR impedance. Lower resistance results in a shorter time to reach the high state, corresponding to a higher force applied to the FSR. The preset GPIO hardware interrupt triggers stereo imaging and records the timestamp when the force applied to the FSR exceeds a certain threshold, indicating that a heel strike has occurred. A quick test to measure the latency of the hardware interruption was conducted by placing a phone running a stopwatch application in front of the stereo cameras during imaging. The result from 15 tests indicates a maximum time interval of 8.67 ms between the imaging start times of the two cameras. Consequently, it is assumed that the two cameras capture images simultaneously. The Universal Asynchronous Receiver Transmitter (UART) communication protocol is used for the wired transfer of the extracted markers' pixel positions between the Raspberry Pi pair on the same wearable. The gait monitoring system also utilizes the User Datagram Protocol (UDP) to wirelessly synchronize time, collect the computed gait parameters from the two wearables, and save the data to the server. In addition, certain redundant designs, such as additional signal input channels, have been incorporated to enhance the system's robustness.

Each portable battery can power the Raspberry Pi for 4-5 hours. The dimensions of the entire wearable are 70 $\times$ 65 $\times$ 32 mm, and it weighs 146.72 grams.

\subsection{Stereo Camera Calibration and Configuration}
\begin{figure}[H]
\centering
\includegraphics[width=3.38in, height=1.36in]{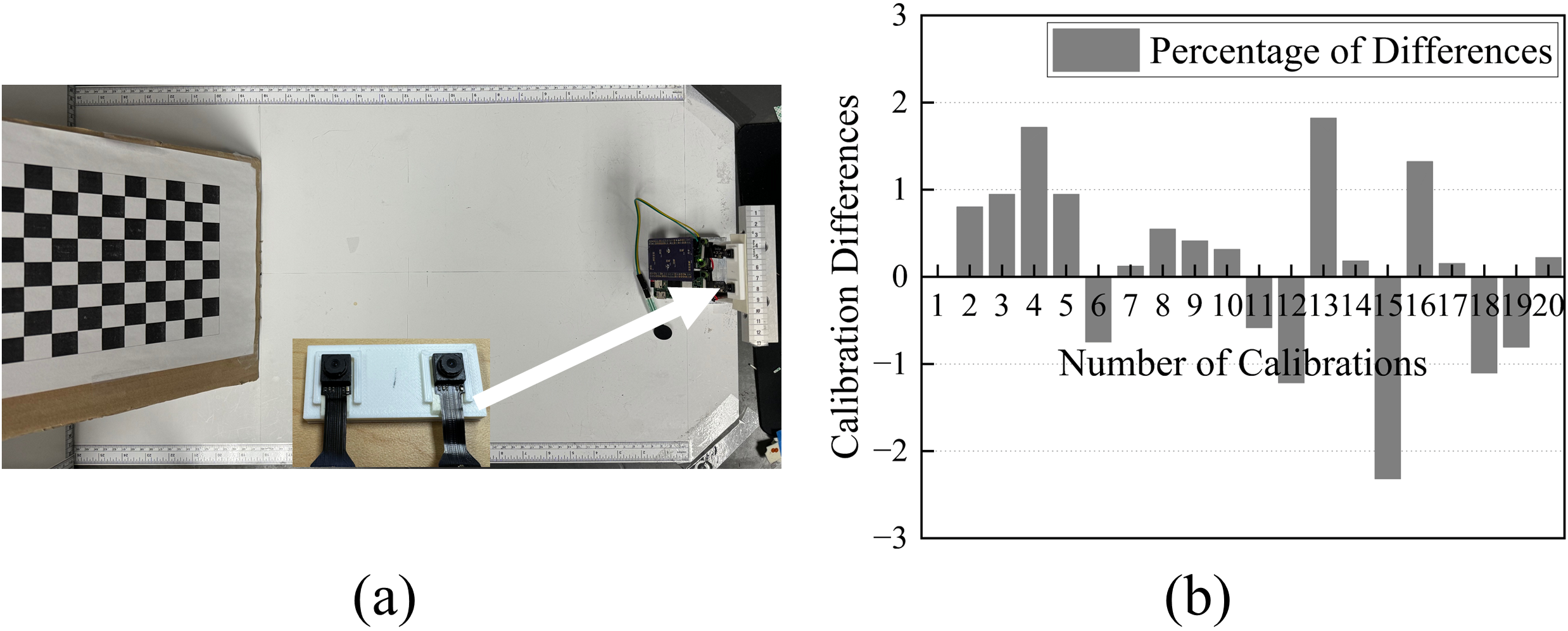}
\caption{(a) Stereo camera calibration setup; the center of the image depicts the holder for the stereo camera. (b) Discrepancies comparison of the x-axis focal length calibration results for the stereo camera with its holder during three months of use, relative to its initial calibration.}
\label{fig_3}
\end{figure}
Camera calibration is essential for accurately measuring gait parameters, especially spatial ones. Fig. 3(a) illustrates the setup for stereo camera calibration. The gait monitoring system, positioned along the center line, captures a series of unique checkerboard stereo images. Given a sufficient number of correspondences $X_i \leftrightarrow x_i$ between 3D checkerboard corner points $X_i$ and 2D image points $x_i$, the camera matrix $P$ can be accurately determined. The camera matrix comprises both the camera intrinsic and extrinsic matrix. The intrinsic matrix correlates to the camera’s internal characteristics, such as focal length, principal points, and distortion coefficients. On the other hand, the extrinsic matrix delineates the positional relationship between two cameras, including rotation and translation.

During the calibration process for the subsequent experiments, we utilized twenty image pairs for calibration and one image pair for a calibration performance assessment. The calibration pattern was positioned at three different locations, specifically at 30 cm, 45 cm, and 65 cm, rotated by either 0 or $\pm$90 degrees, and adjusted to different tilt angles. The reprojection error, which measures calibration accuracy, is defined as the pixel distance between a checkerboard corner key point detected in a calibration image and the corresponding world point projected back into the same image using the calibrated camera matrix. The reprojection error for the experiments described in Section V ranged from approximately 0.2 to 0.45 pixels. The calibration process is conducted using the MATLAB Computer Vision Toolbox.

The stereo camera only requires a one-time calibration as long as the cameras' relative positions remain unchanged. However, in practice, each camera's position may shift due to shaking, vibration, and the deformation of shoes during walking. As shown in the middle of Fig. 3(a), a 3D-printed tough PLA platform is used as the camera holder for the gait monitoring system. Protrusions on the PLA surface surround the stereo cameras to secure their positions. Environmental changes can induce movement in the platform, leading to a simultaneous shift of the entire stereo camera setup while maintaining the relative positioning of the cameras. Fig. 3(b) displays changes in an intrinsic property of the stereo camera, the $x$-axis focal length, over three months of usage to verify the sustained stability achieved by the camera holder. During this period, twenty calibrations were performed, each conducted after an experiment described in Section. V. The results show that, compared to the initial calibration, the calibrated property of the stereo camera did not change significantly (maximum deviation: 2.32\%), indicating that the cameras maintained their positions over an extended period. Consequently, the system does not require continuous camera calibration, thereby eliminating the need for manual intervention and saving time for experts.
\begin{figure}[H]
\centering
\includegraphics[width=3.33in, height=1.82in]{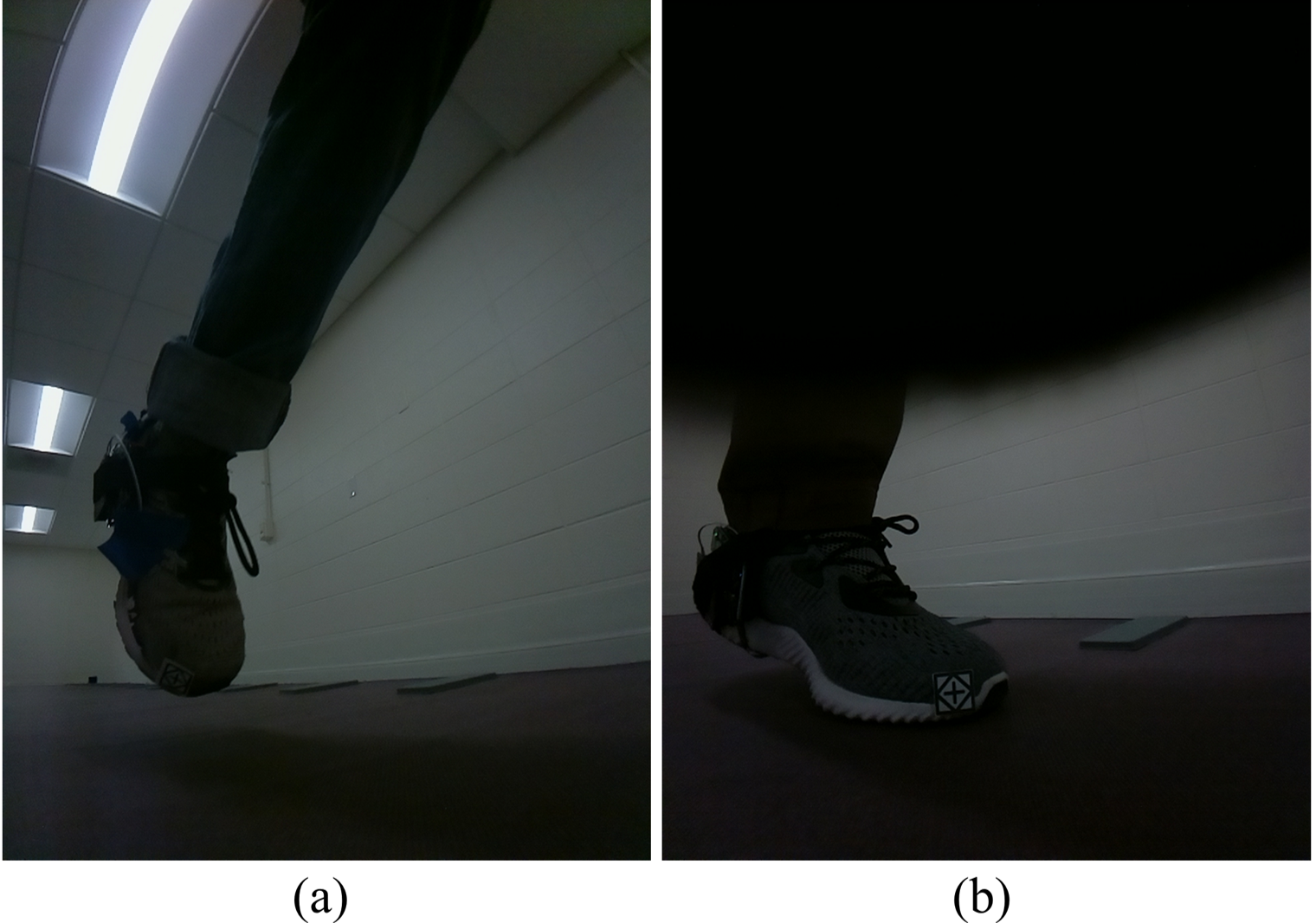}
\caption{(a) An image showing a participant's foot in mid-air during stereo imaging with the FSR placed near the foot tip; (b) Stereo imaging is obstructed when another participant wears long pants during testing.}
\label{fig_4}
\end{figure}
The imaging settings of the stereo camera are critical for enabling rapid image capture and ensuring high-quality image output. A blurred image can lead to inaccuracies in subsequent processing tasks, such as marker detection, resulting in incorrect position identification. A maximum ISO is set to improve the light sensitivity of the stereo camera and to complete exposure quickly by increasing the overall gain. Additionally, a much higher shutter speed has been adopted to reduce exposure time and capture the high-speed motion of the foot. To ensure image consistency during walking, the automatic white balance and exposure compensation are turned off. Meanwhile, the analog and digital gains are fixed. A slightly higher sharpness setting is used for better marker detection. Based on the above settings, the system takes 0.09-0.12 seconds to capture stereo images. Image examples are shown in Fig. 2, Fig. 7(b), and Fig. 12.
\subsection{Installation of the Wearable Device}
\begin{figure*}[b]
\centering
\includegraphics[width=7.13in, height=2.06in]{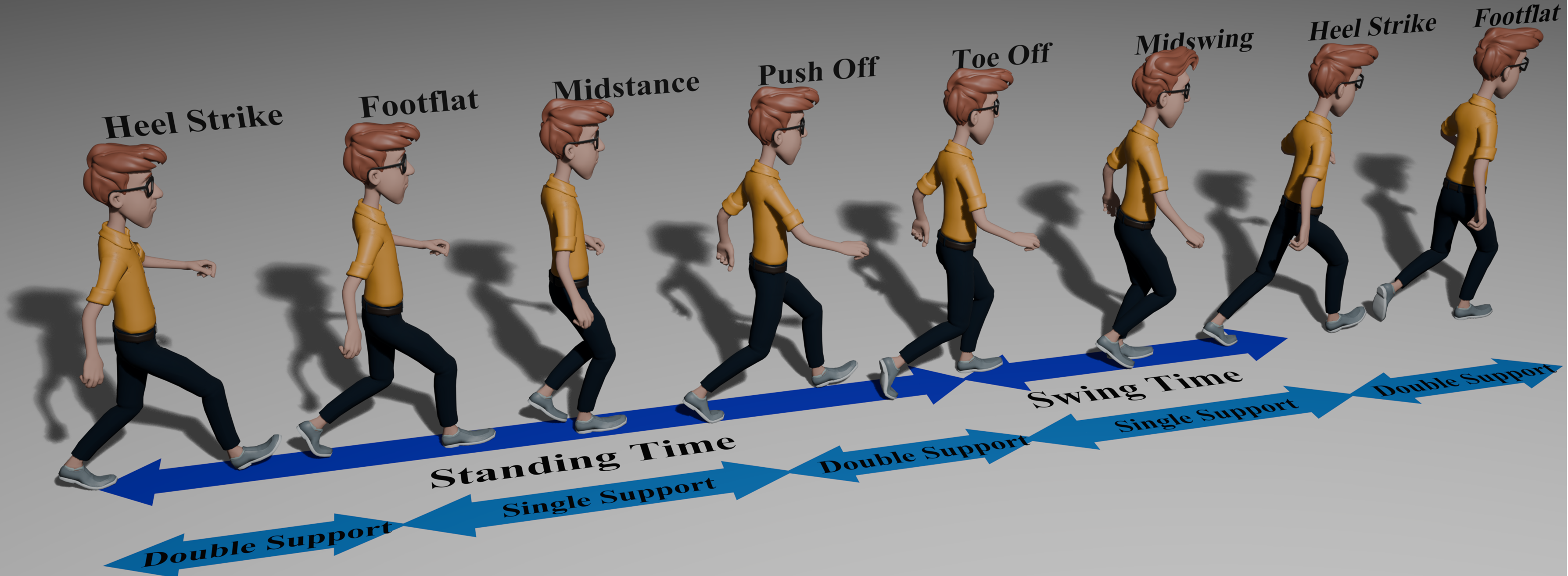}
\caption{Diagram illustrating the walking phases of the right foot and the relationship between temporal parameters. The gait cycle consists of two phases, standing/swing phases, which can be further decomposed into single support and double support time (Free characters from Blender Studio).}
\label{fig_5}
\end{figure*}
The installation of the system on shoes follows two rules: ensuring the accurate measurement of gait parameters and maintaining stability while walking. Since the foot tip is always the last part to leave the ground during walking, it serves as a reliable indicator of the foot's position. Therefore, a self-designed marker was placed on the tip of the shoes. The placement of the FSR beneath the foot is crucial, as it determines the timing of stereo image capture and affects the appearance of the opposite foot within the image. If the FSR is placed too close to the tip, the other foot may be in the air when images are taken, leading to inaccurate measurement of spatial parameters, as shown in Fig. 4(a). Thus, the FSR was placed at the center of the heel. Additionally, compared with our previous work \cite{ieee28}, we simplified the setup by using only one FSR at the foot heel instead of two FSRs at both the tip and heel. This reduced hardware complexity, accelerated computation, and made the device easier to install on the shoes. To fit different shoes with various textures and materials, masking tape was used to cover the shoe's surface, providing a uniform surface. The camera holders were polished with sandpaper to increase surface roughness, enhancing their adherence to the tape. Double-sided tape was then used to attach the platform securely to the masked shoes. Velcro straps can be adjusted to wrap the device on the side of the shoes, ensuring a comfortable fit for most participants. This setup improved the stability of the device and reduced drift during long periods of walking. Note that long pants are not suitable for the system, as they may cover the camera and obstruct marker capture, as shown in Fig. 4(b).

\section{Gait Parameters}
\begin{table}[H]
\caption{Measured Gait Parameters and Their Definitions\label{tab:table1}}
\centering
\begin{tabular}{p{0.84in}|p{2.3in}}
\hline
\hline
\multicolumn{1}{c|}{Gait Parameters} & \multicolumn{1}{c}{Definitions} \\
\hline
Gait Length & Measured along the length of the walkway, between the heel center of the current foot and the tip center of the opposite foot in the previous step\\

Gait Width & Measured vertically along the length of the walkway, between the heel center of the current foot and the tip center of the opposite foot in the previous step\\

Gait Height & Measured height difference between the stereo camera on the current foot and the marker on the opposite foot while walking\\

Stride Length & Measured along the line of progression between the heel centers of two consecutive footprints of the same foot\\

Number of Steps & Number of steps taken to cover a certain distance\\

Cadence & Number of steps taken within a certain period of time\\

Step Time & Time elapsed from the first contact of one foot to the first contact of the opposite foot\\

Stride Time & Time elapsed between the first contacts of two consecutive footfalls of the same foot\\

Swing Time & Time elapsed between the last contact of the current footfall and the first contact of the next footfall on the same foot\\

Standing Time & Time elapsed between the initial contact and the last contact of a single footfall\\

Signal Support T & Time elapsed between the last contact of the opposite footfall and the initial contact of the next footfall of the same foot\\

Double Support T & Time elapsed between heel contact of one footfall to toe-off of the opposite footfall\\

Gait Cycle Time & Time elapsed between the first contacts of two consecutive footfalls of the same foot\\

Ambulation Time & Time elapsed between the first contact of the initial footfall and the last footfall\\

Stride Velocity & Obtained by dividing the stride length by the time\\

Gait Variation & Variation in gait parameters across several steps, which is inversely related to gait stability.\\

Gait Symmetry & The difference in gait parameters between the left leg and the right leg in a single step.\\
\hline
\hline
\end{tabular}
\end{table}
In this section, the definitions of the gait parameters measured by our gait monitoring system, as well as the functions of these parameters, are provided. We further detail the algorithm used to assess temporal gait parameters and explain the distribution of these computational tasks across two Raspberry Pis for parallel processing. Spatial gait parameters will be discussed subsequently in Section IV.
\subsection{17 Measured Gait Parameters}
Fig. 5 shows the model of the walking phase. A full gait cycle consists of two phases: the standing phase and the swing phase. The standing phase includes the heel strike, foot flat, midstance and toe-off phases. The swing phase begins with toe-off, progresses to midswing, and concludes with the next heel strike. Table 1 presents the 17 gait parameters, along with their definitions, that our gait monitoring system is capable of measuring. These gait parameters are categorized into four groups: spatial parameters, temporal parameters, spatiotemporal parameters, and other related parameters. It is important to note that certain gait parameters are interrelated. For example, the single support time of one foot is equal to the swing time of the opposite foot, and the double support time can be determined by subtracting the swing time of one foot from the step time of the opposite foot.

The gait parameters measured by our gait monitoring system can serve as indicators of potential falls or gait abnormalities, revealing underlying changes caused by aging or neurological pathologies \cite{ieee29}. Variations in stride length and gait velocity are associated with declines in executive function \cite{ieee30}\cite{ieee31} and can aid in distinguishing subtypes of mild cognitive impairment, while differences in cadence, swing time, and standing time are related to memory decline \cite{ieee32}. For instance, increased variability in stride time has been identified as an indicator of cognitive decline in older adults, suggesting that this metric could enhance the prediction of dementia, such as Alzheimer's disease \cite{ieee33}. Hypokinetic gait \cite{ieee34} and increased gait variability during attention demanding task, including decreased dual-tasking ability while walking \cite{ieee35}, also appear to be strong markers of decline in gait control for individuals with Parkinson's disease \cite{ieee36}. Compared to healthy older adults, multiple elderly fallers exhibit changes in gait parameters, including slower walking speeds, shorter stride lengths, longer double support times, wider gait widths, and increased variability in stride length and swing time \cite{ieee37}\cite{ieee38}\cite{ieee39}. 
\begin{figure}[H]
\centering
\includegraphics[width=3.49in, height=1.89in]{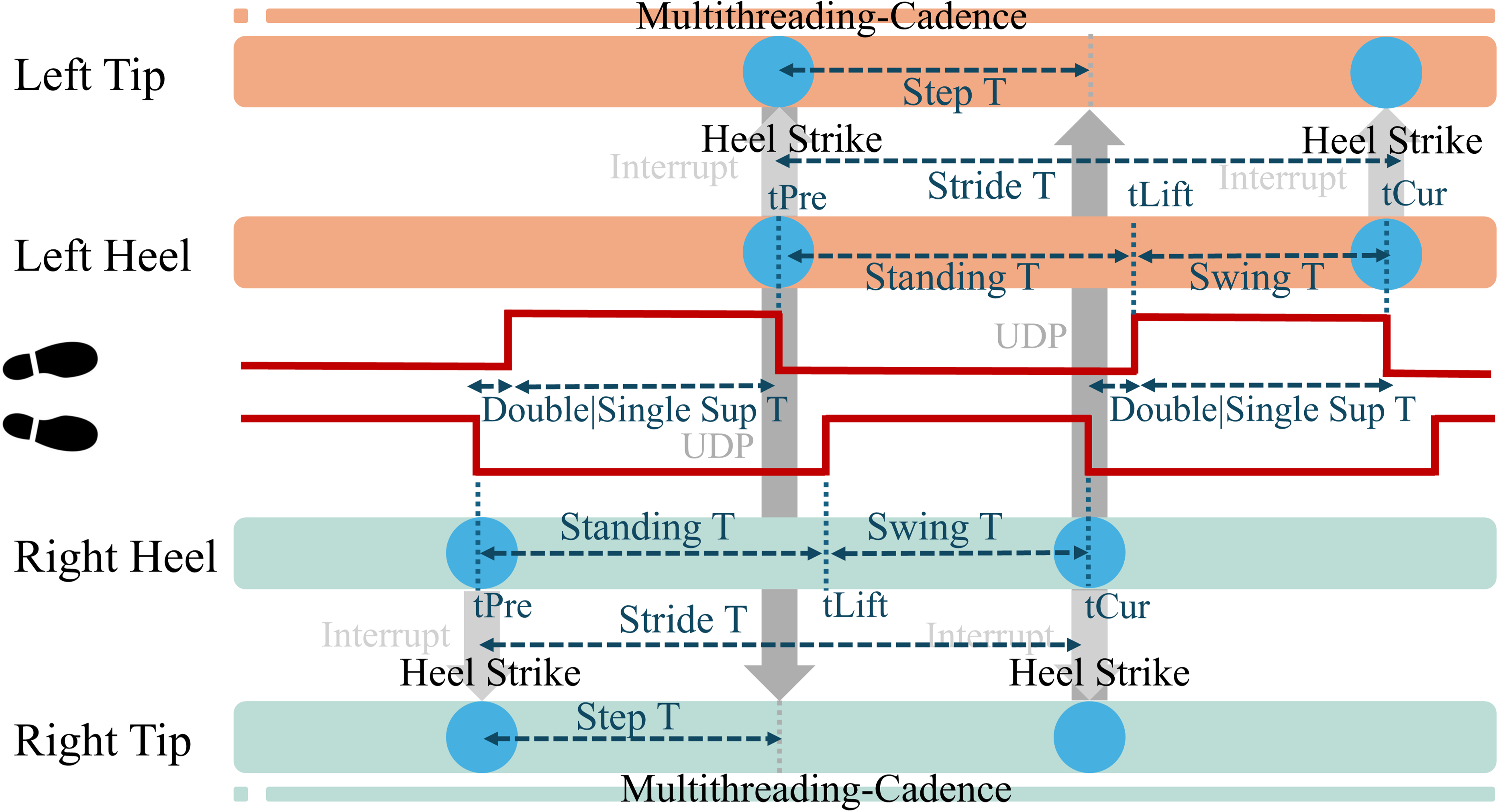}
\caption{Diagram illustrating the algorithm for calculating temporal gait parameters and the implementation of parallel computing.}
\label{fig_6}
\end{figure}

\subsection{Calculation Methodology for Gait Parameters}
Here, we present the algorithms and equations used to measure temporal gait parameters along with other relevant gait metrics. As shown in Fig. 6, the parameter $tPre$ and $tCur$ represent the previous and current step timestamps, respectively. The time difference between these two timestamps is the stride time. $tLift$ is the time when the foot tips are lifted from the ground, splitting the stride time into standing time and swing time. Since only a single FSR is positioned at the foot heel, the foot lift time is determined after a time count of 100 on the FSR following the heel strike. Subsequently, a compensation factor is applied to all collected measurements. Step time is measured by calculating the time difference between the heel strikes of the left and right foot, using timestamps data synchronized via UDP. The double support time can be determined by subtracting the swing time from the step time. The single support time of one foot corresponds to the swing time of the opposite foot. Cadence and step counts are monitored through a separate parallel thread. Ambulation time can be accumulated according to the step time and step counts described above.

The accuracy of measured temporal gait parameters also depends on the gait monitoring system's time resolution and latency. The Raspberry Pi can distinguish time intervals in microseconds. The average time resolution from 15 experiments, where two timestamps are returned immediately, is 7.20473E-06 seconds. The round-trip UDP latency, i.e., the time intervals for the server to send a message (48-bit) and receive a confirmation message of the same length from the client, averaged over 20 experiments, is 0.004169 seconds.

Additionally, the system provides two statistical gait parameters, described below. The variation of gait parameters across several continuous steps is defined as,
\begin{equation}
s = \sqrt{\frac{1}{N-1} \sum_{i=1}^{N} \left(x_i - \bar{x}\right)^2}, \ \text{where} \ \bar{x} = \frac{1}{N} \sum_{i=1}^{N} x_i
\end{equation} 
\begin{equation}
\%CV = \frac{s}{\bar{x}} \times 100
\end{equation}
where $x_i$ represents each gait parameter in a sequence, $\bar{x}$ and $s$ are the mean and sampled standard deviation, and $\%CV$ is the percentage of coefficient of variance. Gait symmetry is calculated as follows,
\begin{equation}
\% \text{Sym} = 2 \times \left| \frac{x_{\text{right}} - x_{\text{left}}}{x_{\text{right}} + x_{\text{left}}} \right| \times 100
\end{equation}
where $x_{\text{right}}$ and $x_{\text{left}}$ denote the gait parameters from the right and left foot, respectively, and $\%Sym$ is the percentage of gait symmetry. Given the extensive range of gait parameters the system can measure, only gait length and stride velocity are considered in this work for evaluating the gait variation and symmetry, as these two parameters are currently prioritized in clinical diagnoses.

\section{Software Design}
\subsection{Marker Design and Detection Algorithm}
Accurate detection of marker positions is essential for calculating spatial gait parameters, as it enables precise determination of shoe positions. A custom-designed marker was developed to integrate with the detection algorithm. As shown in Fig. 7(a), it features a rare and unique pattern that can be easily distinguished from common everyday objects, allowing for reliable detection even when partially obscured by image borders. Additionally, the square shape of the marker simplifies annotation during training, and a cross mark is incorporated to facilitate quick identification of the marker's center. The dimension of the marker is 3 $\times$ 3 cm.

We use YOLO (You Only Look Once) \cite{ieee40} as our primary method for marker detection due to its advantages in localizing marker positions under complex lighting conditions. YOLO is a real-time object detection algorithm that formulates object detection as a single regression problem. It divides an image into grids, with each cell responsible for predicting bounding boxes and associated class probabilities. The algorithm utilizes residual blocks for feature extraction, bounding box regression for accurate localization, Intersection Over Union (IoU) for evaluating overlapping boxes, and Non-Maximum Suppression to remove redundant boxes.

The training images were collected independently and prior to the experiments. A participant walked with a single gait monitoring device attached to the right foot, completing a total of 838 steps. Only one image from each pair of stereo images was used. During certain steps, the participant intentionally altered gait to obscure the marker, generating negative samples. This approach enhanced the model's ability to correctly identify the absence of the marker, thereby reducing false positives. The collected images were augmented under the following conditions: with rotations between $\pm$20 degrees, brightness adjustments between $\pm$20\%, exposure changes between $\pm$10\%, and a uniformly applied Gaussian blur of up to 2 pixels. In total, there were 1,758 images for training, 169 for validation, and 83 for testing. The training process took 15.41 hours over 50 epochs on a Nvidia GeForce RTX 4070 Ti. The average inference time was 41 ms per image. The testing results are shown in Fig. 7(b).
\begin{figure}[H]
\centering
\includegraphics[width=3.47in, height=1.48in]{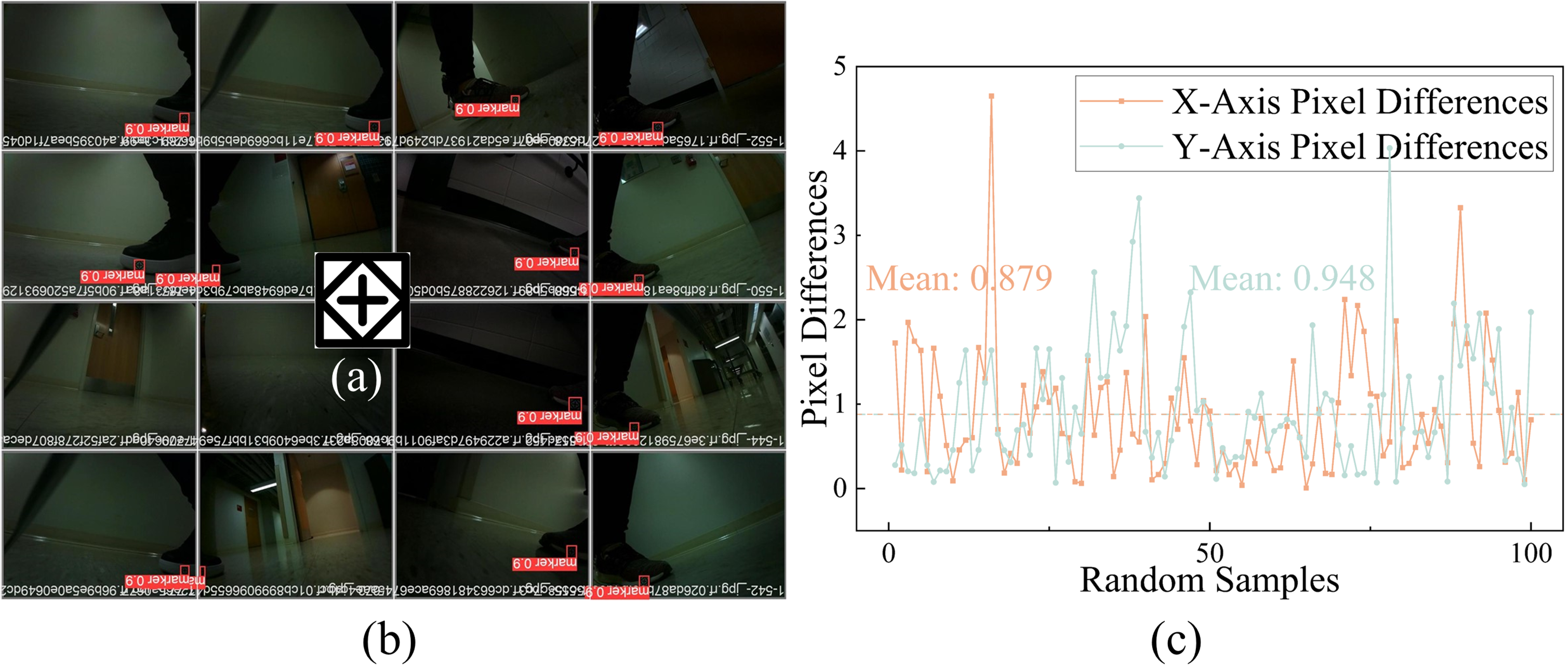}
\caption{(a) A custom-designed marker used for foot positioning; (b) Results of marker detection using the trained YOLO model; (c) Pixel differences in marker positions between manual labeling and YOLO detection results.}
\label{fig_7}
\end{figure}
The mean Average Precision (mAP) at an IoU threshold of 50-95\%, i.e., the precision of marker prediction where the predicted bounding box overlaps with the ground truth box by at least 50\%, is 94.7\%. Given its square shape, the four corner pixel positions are extracted from the bounding box to compute the center point, which is designated as the marker’s position. As shown in Fig. 7(c), we randomly sampled 100 images from subsequent experiments and compared the pixel differences in marker positions between manual labeling and YOLO detection results. The differences are 0.879 pixels along the x-axis and 0.948 pixels along the y-axis, respectively, indicating a high level of accuracy. In practice, we observed that the markers were undetected in five stereo images during later experiments due to significant tilting. 

\subsection{Spatial Parameters Calculation}
Triangulation is utilized to determine the 3D positions of the markers. The marker pixel positions from the stereo image pair are combined to compute the marker's 3D reference positions, with one camera chosen as the reference camera. The reference positions are then used to compute the world coordinate spatial gait parameters.

The mapping from the marker's 3D world coordinates to 2D image points can be described as follows \cite{ieee41},
\begin{equation}
X_{\text{cam}} = 
\begin{bmatrix}
R & -R\tilde{C} \\
0 & 1 
\end{bmatrix}
X
\end{equation}
\begin{equation}
x = K \begin{bmatrix} I & 0 \end{bmatrix} X_{\text{cam}} 
\end{equation}
By combining the above two equations, we get the following,
\begin{equation}
x = KR \begin{bmatrix} I & -\tilde{C} \end{bmatrix} X 
\end{equation}
where $x \in \mathbb{R}^{3 \times 1}$, $X_{\text{cam}} \in \mathbb{R}^{4 \times 1}$, and $X \in \mathbb{R}^{4 \times 1}$ represent the homogeneous image, camera, and world coordinate marker points, respectively. $K$ and $R$ are the camera calibration matrix and rotation matrix, respectively, each with dimensions of $\mathbb{R}^{3 \times 3}$. $\tilde{C}$ is the $\mathbb{R}^{3 \times 1}$ camera center in world coordinate. $I$ is the identity matrix. By making the camera center implicit, where $\tilde{X}_{\text{cam}} = R \tilde{X} + t$, (6) can be simplified as below,
\begin{equation}
x = PX = K \begin{bmatrix} R & t \end{bmatrix} X 
\end{equation}
where $P$ is the projection matrix that includes the internal matrix $K$ and external matrix $R$ and $t$. $t$ is the product of $-R\tilde{C}$, which is referred to as the translation matrix.

The camera internal matrix $K$ has the form:
\begin{equation}
K = 
\begin{bmatrix}
fm_x & s & x_0 \\
0 & fm_y & y_0 \\
0 & 0 & 1 
\end{bmatrix}
\end{equation}
where $fm_x$ and $fm_y$ represent the focal lengths of the camera in terms of pixel dimensions along the $x$ and $y$ axes, respectively, $x_0$ and $y_0$ are the principal point offsets. $s$ denotes skewness, which occurs when the axes of the image are not perpendicular, causing square content within a pixel to appear distorted, taking on a parallelogram shape. However, in modern cameras, skewness is generally negligible, so $s=0$.

As one camera in a stereo camera setup is designated as the reference camera, its projection matrix is $P_1 = K_1 \begin{bmatrix} I & 0 \end{bmatrix}$. The internal camera matrix $K_1$ and the projection matrix of the second camera $P_2 = K_2 \begin{bmatrix} R_2 & t_2 \end{bmatrix}$ can be obtained through the camera calibration described in Section II. B.

In stereo imaging, the marker's position in world coordinates is first transformed into camera coordinates, followed by mapping the camera coordinates to image coordinates. Conversely, the 3D position of the marker within a scene can be computed using projection matrices of each camera and the marker's projections onto the stereo images. If $o_1 = \begin{pmatrix} x_1, y_1, 1 \end{pmatrix}^T$ and $o_2 = \begin{pmatrix} x_2, y_2, 1 \end{pmatrix}^T$ are two pixel points on stereo images representing the projections of the same 3D marker point $X = \begin{pmatrix} X_1, Y_1, Z_1, 1 \end{pmatrix}^T$, then $X$ can be reconstructed using $P_1$, $P_2$, $o_1$, and $o_2$.

For each image pair, we have $o_1 = P_1X$ and $o_2 = P_2X$. These two equations can be combined into the form $AX=0$, which is a linear equation in $X$. The homogeneous scale factor is eliminated by a cross product of $o_i \times (P_i X) = 0, i = 1, 2$, then $AX=0$ can be composed with,
\begin{equation}
A = 
\begin{bmatrix}
x_1 P_1^{3T} - P_1^{1T} \\
y_1 P_1^{3T} - P_1^{2T} \\
x_2 P_2^{3T} - P_2^{1T} \\
y_2 P_2^{3T} - P_2^{2T}
\end{bmatrix} 
\end{equation}
where \( P_j^{iT} \) are the \( i \)th row of \( P_j \), \( i = 1, 2, 3 \), \( j = 1, 2 \). The least squares solution of \(AX=0\) subject to \(\|X\| = 1\), i.e., the 3D position values, is given by the last column of $V$, where $A = U \Sigma V^T$ is the singular value decomposition of $A$ \cite{ieee41}.

The calculated 3D positions are in the reference camera coordinate. Therefore, a coordinate transformation to the ground coordinate is necessary to extract the spatial gait parameters. Since the gait monitoring system is mounted on the shoes parallel to the ground, a point $x$ in the reference camera coordinates can be transferred to $x'$ in the ground coordinates through a transformation matrix, as shown below,
\begin{equation}
x' = 
\begin{bmatrix}
\cos\theta & -\sin\theta \\
\sin\theta & \cos\theta
\end{bmatrix}
x 
\end{equation}
\begin{figure}[H]
\centering
\includegraphics[width=3.47in, height=3.93in]{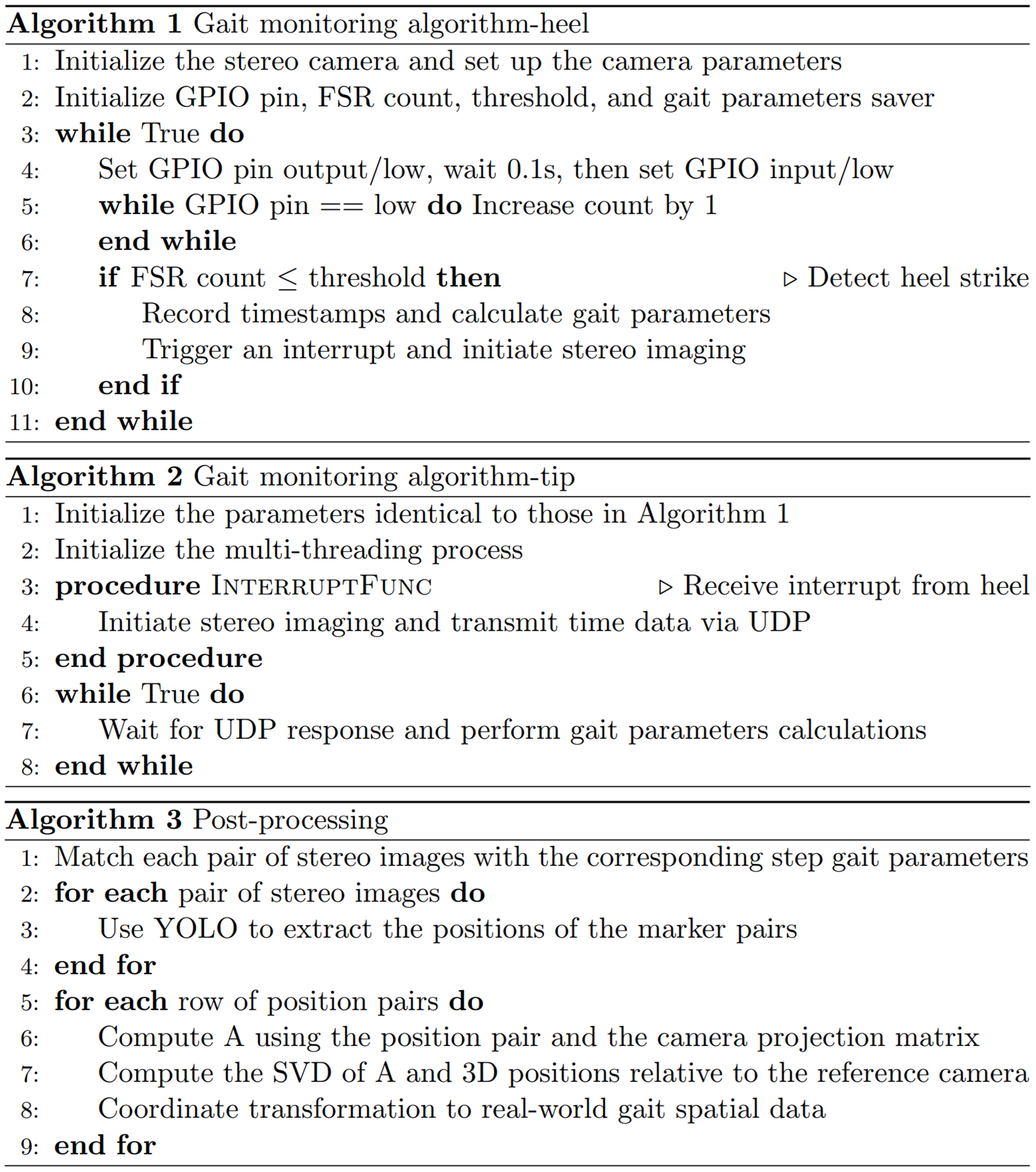}
\caption{The pseudocode for operating the designed gait monitoring system involves data capture, temporal data measurement, and image post-processing to extract spatial gait parameters.}
\label{fig_8}
\end{figure}
\noindent where $\theta$ represents the angle between the camera's line of sight and the walking direction, which can be measured during the installation of the device. As described in Section III. A., the gait height refers to the height difference between the camera on one wearable and the marker on another. The initial height difference between the two wearables, when both feet are in the foot-flat phase, is measured in advance. Subsequent gait height calculations use the measured height values, incorporating the initial height difference to obtain the final gait height. Additionally, the stride length is calculated as the sum of two consecutive gait lengths plus the length of the foot.

\subsection{Pseudocode for Gait Monitoring}
As depicted in Fig. 8, the algorithms for the gait monitoring system are divided into three components: the monitoring algorithm for the gait heel and tip devices in each wearable, and the post-processing stage for image analysis and gait parameters. For better notation, the two Raspberry Pis in each wearable are referred to as the gait heel and the gait tip device.

The real-time stereo image capture and temporal gait parameter calculations are performed using the gait heel and tip devices. The gait heel device is primarily responsible for detecting heel strike events, which trigger a hardware interrupt. This signal allows the gait tip device to simultaneously capture images and count steps using multithreading. The temporal data calculations are distributed across the two devices to enable parallel processing. Post-processing includes YOLO inference for marker detection and the calculation of spatial gait parameters. Parameters such as gait speed, variation, and symmetry are computed by integrating all spatial and temporal parameters.

\subsection{Hardware Potential for Integration with LLMs}
\begin{figure}[H]
\centering
\includegraphics[width=3.47in, height=1.37in]{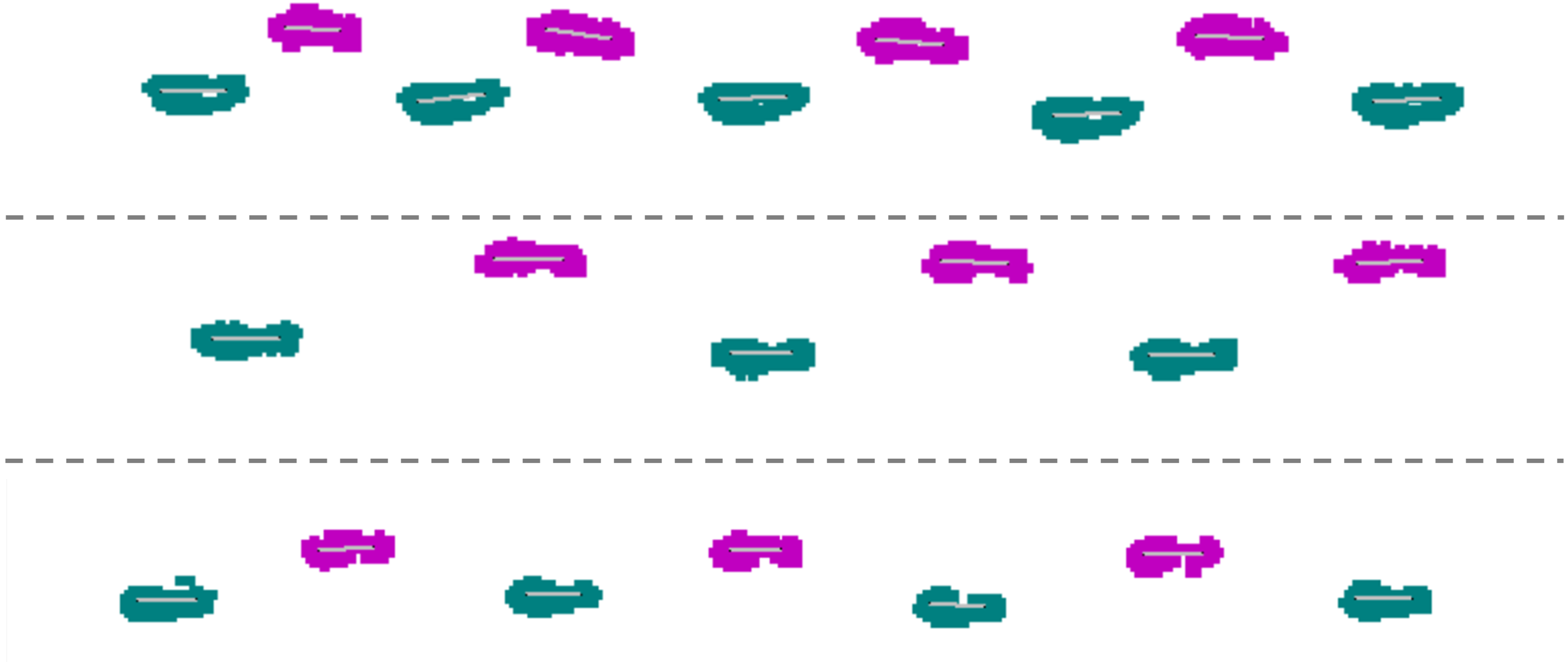}
\caption{The visualized gait patterns of three participants.}
\label{fig_9}
\end{figure}
Given the growing interest in disease diagnosis using LLMs, the gait monitoring system we developed could serve as a valuable hardware tool for collecting long sequences of gait data, which can be effectively utilized in LLMs. As shown in Fig. 9, the visualized gait patterns of the three participants demonstrate that gait is a unique sequence biometric characteristic. It can be used for person identification and for detecting subtle variations that may signal potential gait-related disorders. As described later in Section VI, our hardware demonstrates the capability to measure 17 gait parameters with high accuracy and minimal signal drift during extended walking. This approach allows us to collect a high-quality, enriched gait dataset that can be integrated with LLMs, offering advantages over traditional technologies like gait mats, which are limited to measuring 5-7 steps per round. We validate this capability through person identification using the data collected from our system.
\begin{figure}[!t]
\centering
\includegraphics[width=3.35in, height=1.6in]{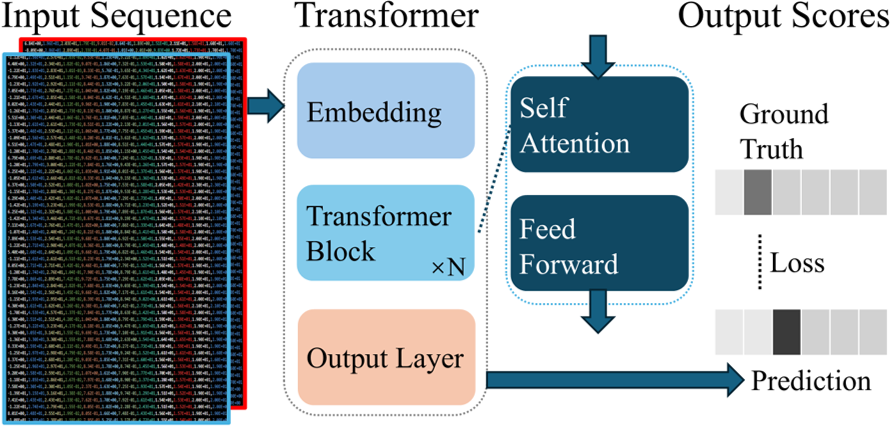}
\caption{The diagram illustrates the Transformer architecture, including an input sequence of gait features, word and position embeddings, Transformer blocks, residual skip connections, and the encoder-decoder structure.}
\label{fig_10}
\end{figure}

The Transformer architecture is applied to validate the effectiveness of our developed hardware in distinguishing individual participants, given its central role in LLMs. As a widely used tool in sequence-related tasks, the Transformer architecture is particularly well-suited for processing long-sequence gait data. As shown in Fig. 10, the typical Transformer block comprises a self-attention layer, layer normalization, a feed-forward layer, followed by an additional layer of normalization. Its self-attention mechanism significantly enhances identification accuracy by establishing strong connections between elements within a sequence, thereby improving the comprehension of the overall context. Moreover, this mechanism avoids the inefficiencies associated with sequential processing, as observed in Long Short-Term Memory, leading to better time performance during both training and inference phases.

We trained a Transformer model to classify input sequences, with each label corresponding to a specific participant. Given the limited size of our dataset, we reduced the number of encoder and decoder layers to mitigate overfitting and achieve faster training times. As detailed later in Section V, each walking cycle comprised approximately 70 steps of gait data. To capture distinct and continuous gait features, the entire sequence was segmented into sub-sequences of length 128 for training. When a sequence was shorter than 128 steps, zero padding was applied. This segmentation approach also facilitated faster convergence by reducing the number of parameters. We implemented random dropout in the network during training to enhance robustness against observational noise.

\section{Experiment}
This section provides a detailed description of the experiments conducted to evaluate the performance of the designed gait monitoring system. The experiments include both short- and long-distance walking tests to assess the system's accuracy and drift, along with extended data collection to analyze walking patterns over a prolonged period. The performance of the system in the first two experiments was evaluated using a gait mat as the reference standard.

Six healthy participants were recruited for the above three experiments, all of whom provided informed consent to participate in the research. Among the participants, five were male, and one was female.

\subsection{Short-Distance Walking Test on the Gait Mat}
\begin{figure}[!b]
    \centering
    \includegraphics[width=3.47in, height=1.18in]{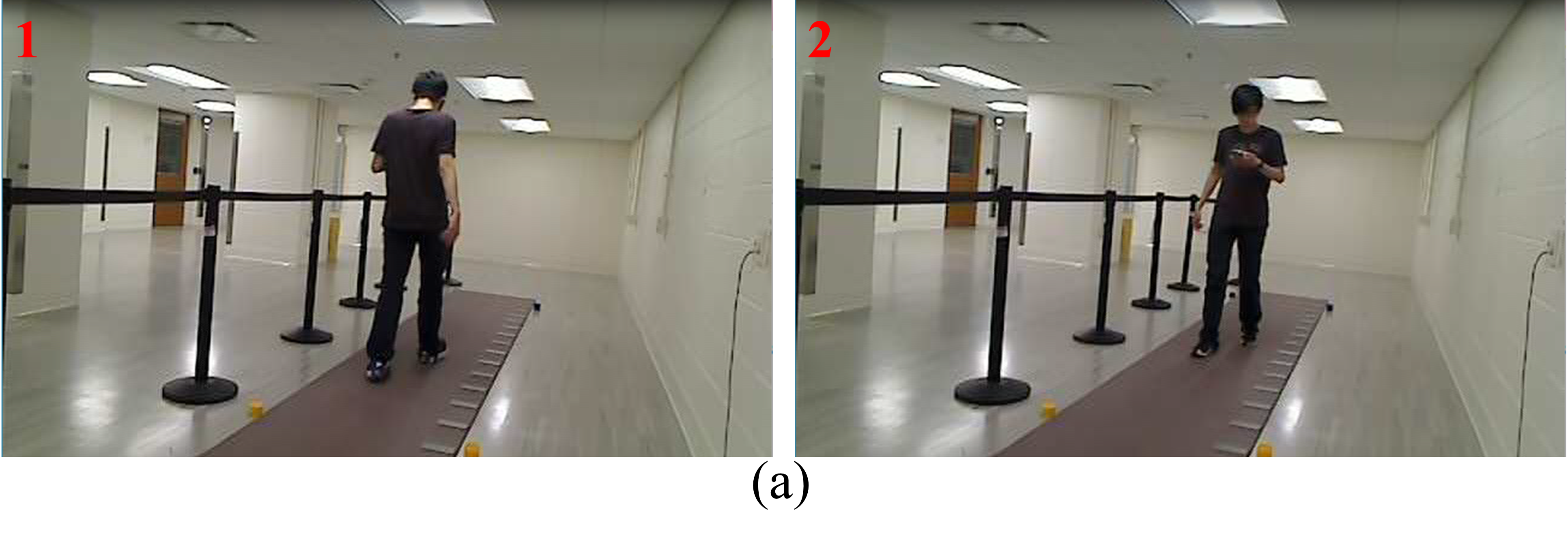}
    \includegraphics[width=3.47in, height=1.02in]{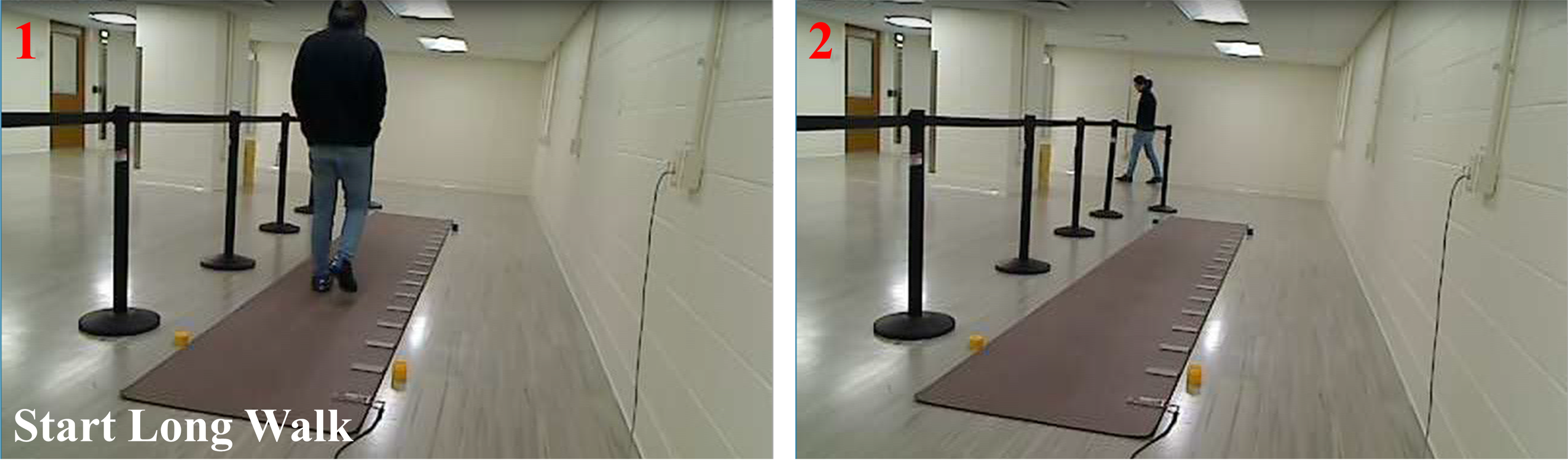}
    \includegraphics[width=3.47in, height=1.18in]{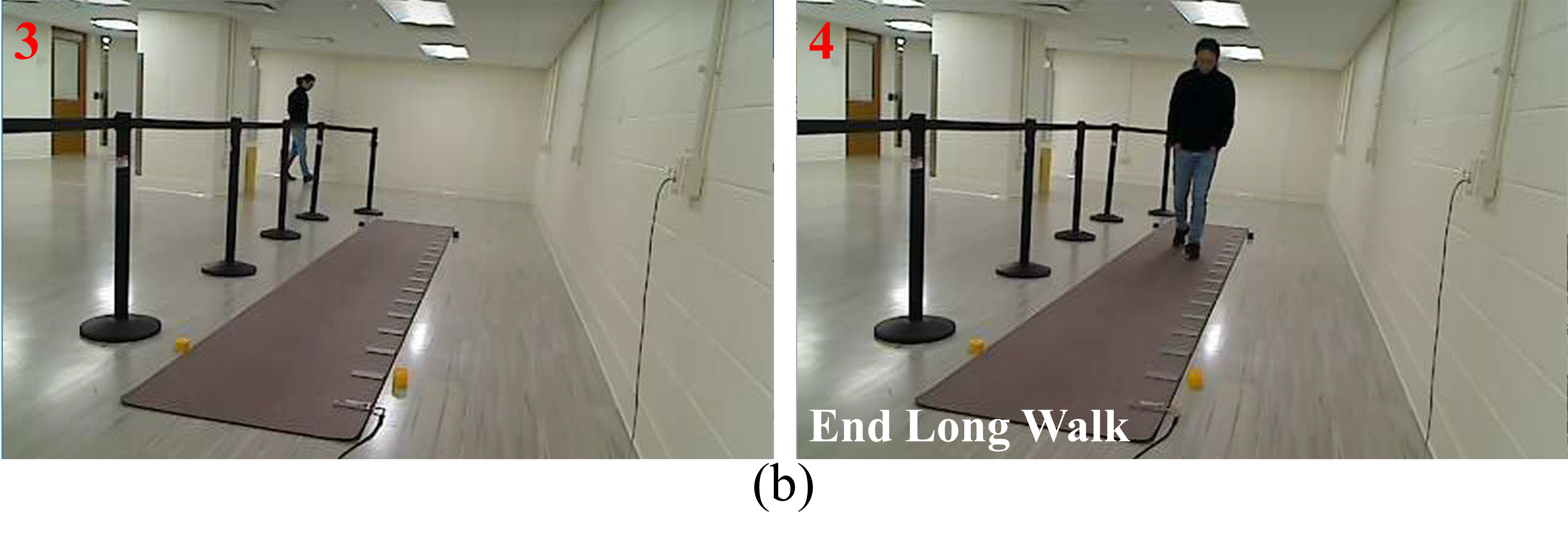}
    \includegraphics[width=3.47in, height=1.4in]{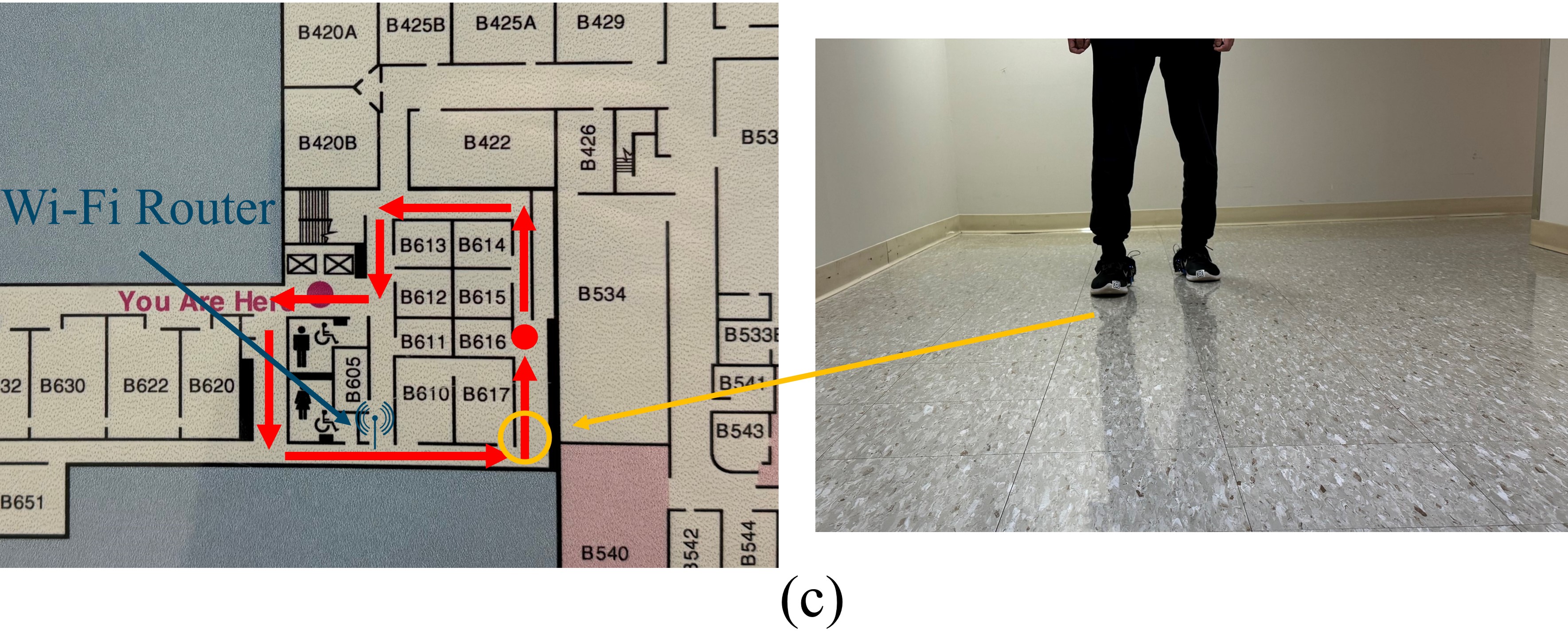}
    \caption{(a) A participant walked back and forth on the gait mat while wearing the gait monitoring system and performing dual tasks; (b) A participant began and ended the long walk task on the gait mat, with approximately 120 steps taken during the middle portion of the task; (c) The image illustrates the walking route used for collecting sequence gait data, with the red dot indicating the start and end point, and the Wi-Fi router positioned in the middle of the route.}
    \label{fig_11}
\end{figure}
As illustrated in Fig. 11(a), participants were instructed to walk back and forth across the gait mat eight times while wearing the gait monitoring system. During these trials, participants engaged in dual tasks, such as checking emails, browsing social media, and texting, to divert their attention from the experiment, thereby helping to preserve their natural gait patterns. A Wi-Fi router positioned next to the gait mat facilitated data transfer between the system and the PC. The length and width of the gait mat are 700 cm and 90 cm, respectively, with most participants taking 6-8 steps to traverse it, except for one participant who took 9-10 steps. A total of 710 steps were collected. We then compared the gait parameters obtained from the gait mat with the data measured by our system to evaluate the performance of the designed system.

However, since we use the gait mat as a comparator, there are certain differences between our system and the gait mat that will be adjusted accordingly. For instance, the definitions of some parameters differ between the two systems. Specifically, the gait mat algorithm defines gait width as the direct distance between heel to heel, which is almost equivalent to gait length when the step is "huge and narrow." In contrast, we define gait width as the vertical distance along the walkway between the tip and heel. The discrepancy is accounted for during the calculation.

Meanwhile, the gait mat cannot measure gait height values. Therefore, to capture the height, participants were instructed to lift their feet to their natural gait height eight times for each foot while walking. The height was marked on the wall to establish the ground truth, which was subsequently compared with the measured values obtained from the gait monitoring system.

\subsection{Long-Distance Walk Test for Drift Assessment}
As illustrated in Fig. 11(b), the test was divided into four phases. Initially, the participant walked forward on the gait mat. Following this, they proceeded to walk off the gait mat to commence the long-distance walk. After that, the participant returned to the gait mat. Finally, the participant walked on the gait mat once more to conclude the test. The long walk was conducted indoors, where participants walked approximately 120 steps, following the room's shape and making three turns. Participants were instructed to maintain their natural gait throughout the walk and completed this walking task three times. A total of 262 steps were recorded on the gait mat. To evaluate the system's drift during long walks, we compared the accuracy of the gait parameters measured by the gait monitoring system between the initial and final walks on the gait mat.

\subsection{Long-Distance Walk Data Collection}
\begin{figure}[!b]
\centering
\includegraphics[width=3.46in, height=1.11in]{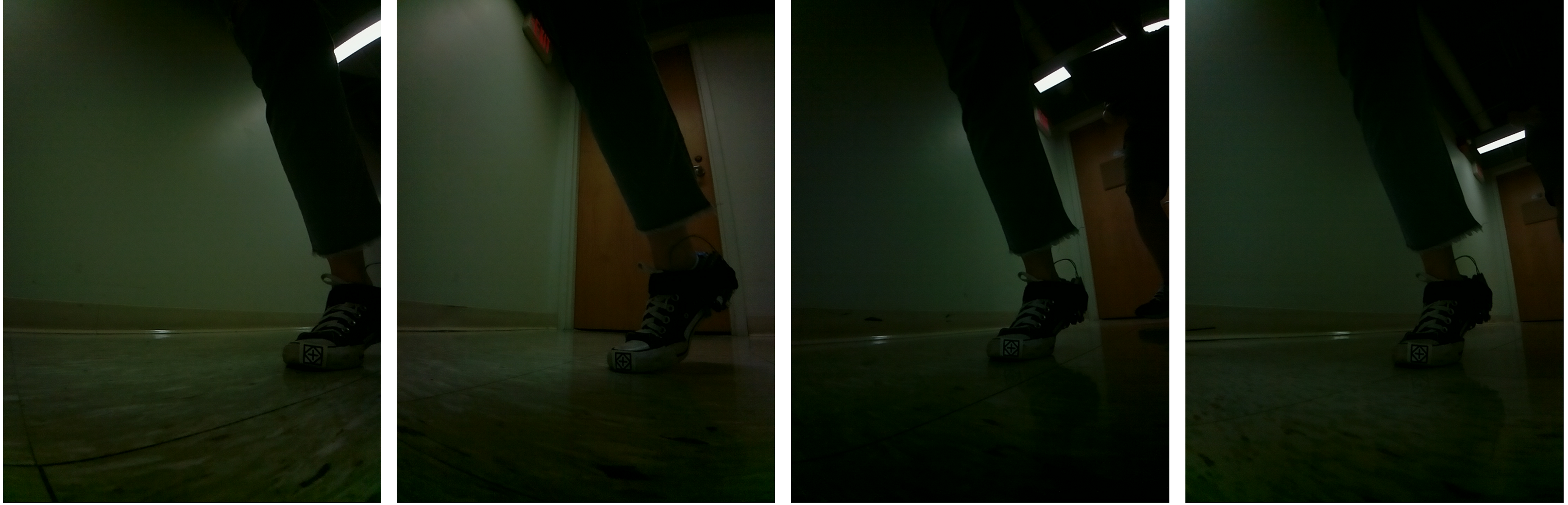}
\caption{Continuous stereo images captured at a corner turn.}
\label{fig_12}
\end{figure}
Fig. 11(c) illustrates the route of the walking trial designed to collect sequential walk data. The red dot marked the start and end points, while the Wi-Fi router was positioned at the center bottom of the route. A walking cycle required approximately 120 steps to complete, depending on the participant and walking circumstances. Participants were instructed to walk around 20 cycles (20, 16, 19, 18, 20, 20) to collect sequential gait data. A total of 14,363 sequential steps, corresponding to 244,171 features, were collected. Additionally, the gait monitoring system was capable of continuously tracking corner turns, as displayed in Fig. 12. Each trial included six corner turns, and the manner in which participants navigated these corners could serve as a valuable feature during training.

\section{Result}
\subsection{Accuracy of the 17 Measured Gait Parameters}
Fig. 13 illustrates the average absolute error for the 17 gait parameters measured by the gait monitoring system compared to the gait mat. The final table provides the accuracy of each gait parameter. Overall, the results demonstrate that our system exhibits high accuracy and precision.

Gait symmetry is not compared with the gait mat, as the mat does not provide this measurement. The measured gait symmetry in the gait monitoring system for six participants, in terms of gait length and velocity, is (2.68\%, 2.31\%), (5.23\%, 8.69\%), (5.00\%, 5.17\%), (4.84\%,4.69\%), (5.61\%, 4.46\%), and (7.53\%, 8.60\%). These values align with expectations, as all participants are healthy individuals; thus, the symmetry, i.e., the differences in gait parameters between left and right footsteps, should be small.

Several findings from the figures are summarized below. Fig. 13(a)-(c) demonstrates that the absolute error for all three-dimensional gait parameters is within 1 cm. However, the stride length exhibits a larger error. This discrepancy arises from the accumulation of errors, given that stride length is calculated as the sum of two consecutive gait lengths and the length of the foot. Besides, the foot length of each step varies on the gait mat due to its resolution and the pressure points of the participants' shoes, introducing additional errors in the stride length comparison. All the temporal gait parameters in Fig. 13(e)-(i), including step time and stride time, exhibit lower absolute error, which can be attributed to the use of precise algorithms, the high time resolution provided by the system, and the correct placement of the device on the foot. It is important to note that the number of data points for temporal gait parameters is lower than the step counts, as these parameters reflect the relationships between two consecutive steps; thus, the first step of each trial is not included. Each subfigure in Fig. 13(j)-(l) presents results for two gait parameters. Note that some discrepancies in the step counts and cadence measurements occur, which can be attributed to the FSR. Occasionally, the FSR triggers the stereo imaging twice for one step due to aging of FSR or adhesion to the bottom of the shoes. Stride velocity is calculated for each step by utilizing the corresponding stride length and time. This measurement also inherits a larger absolute error from the stride length. The overall variation in gait length and velocity is minimal, indicating that the gait monitoring system is capable of accurately tracking trends and changes in gait parameters.

\subsection{Device Measurement Drift During Long-Distance Walks}
\begin{figure*}[t]
    \centering
    \includegraphics[width=7.17in, height=1.72in]{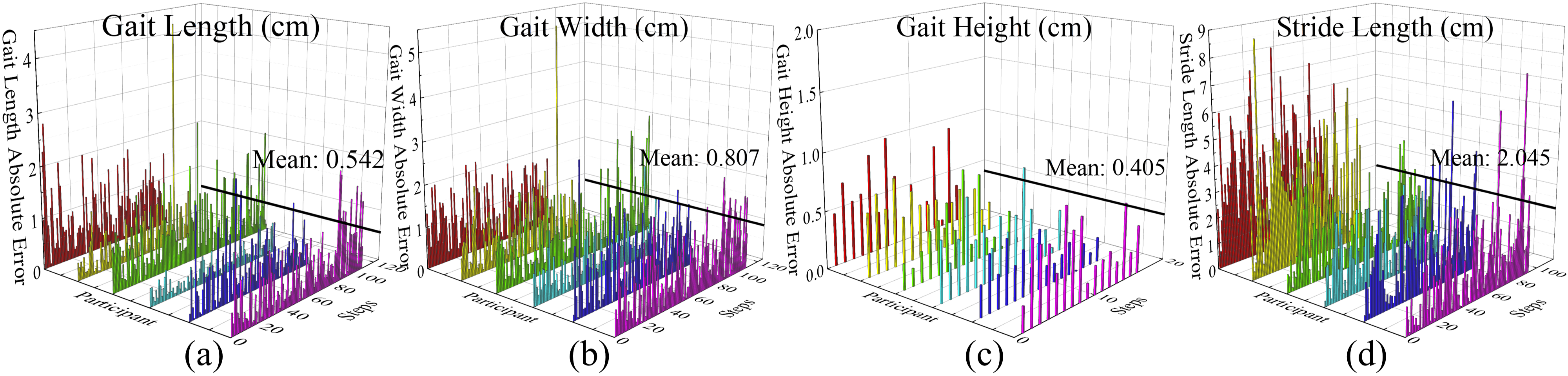}
    \includegraphics[width=7.17in, height=1.72in]{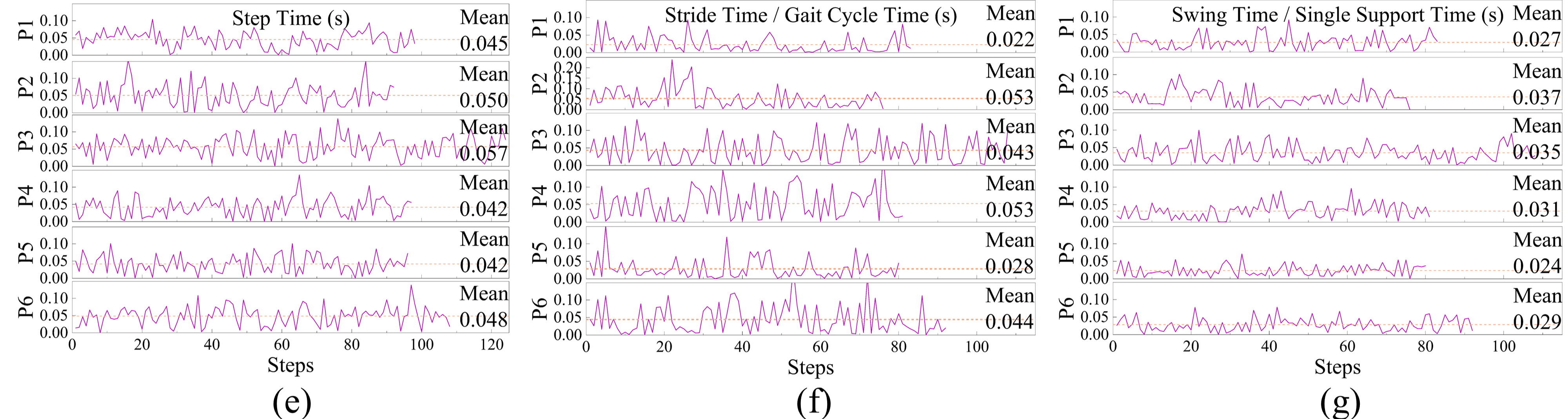}
    \includegraphics[width=7.17in, height=1.72in]{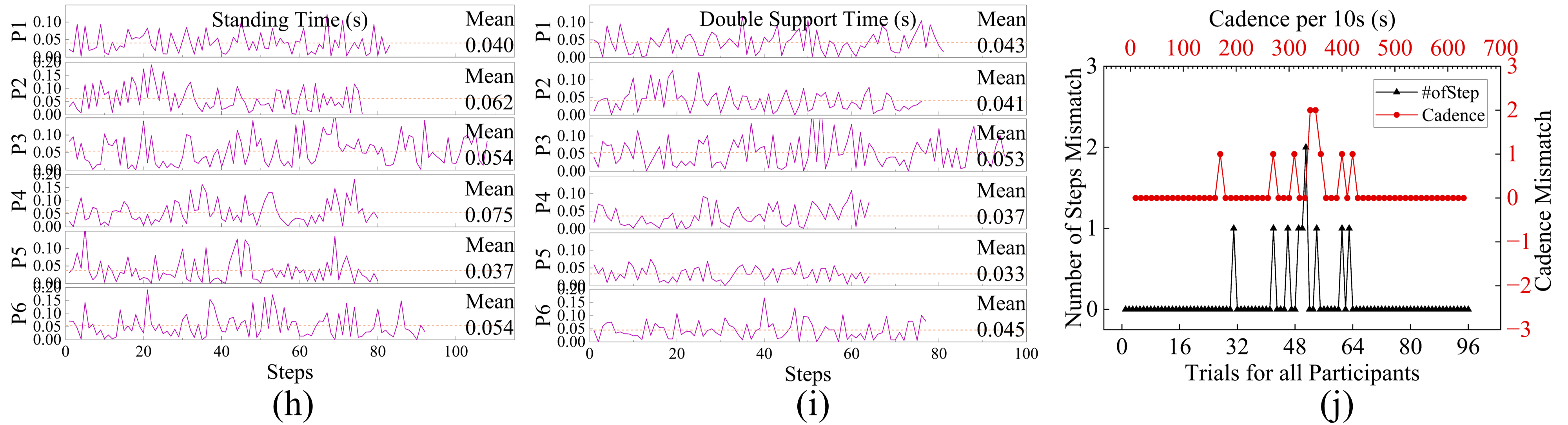}
    \includegraphics[width=7.17in, height=1.72in]{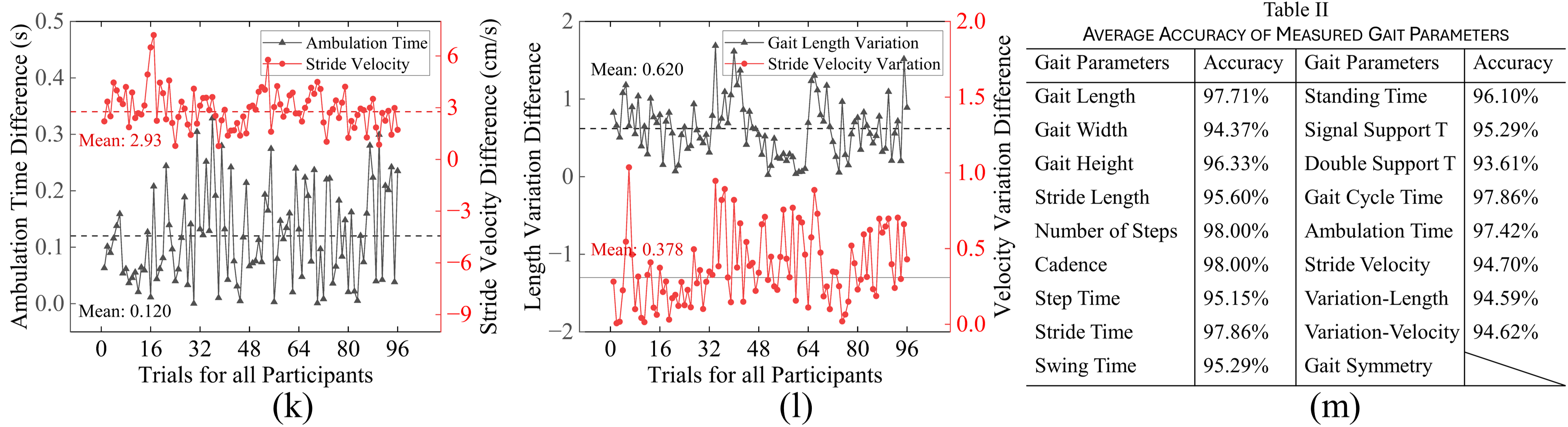}
    \caption{The accuracy of measured gait parameters for six participants compared to the gait mat. (a)-(d) represent the absolute errors in spatial parameters: gait length, gait width, gait height, and stride length; (e)-(i) show the absolute errors in temporal parameters: step time, stride time, gait cycle time, swing time, single support time, standing time, and double support time; (j)-(l) illustrate the absolute errors in the number of steps, cadence, ambulation time, stride velocity, and gait variation; (m) displays a table summarizing the accuracy for all gait parameters.}
    \label{fig_13}
\end{figure*}

As shown in Fig. 14(a), the difference in absolute error across 17 gait parameters (both gait variation and symmetry include two measures, length, and velocity) between the start and end of the long walk, averaged over three trials, is less than 1.5. Fig. 14(b) displays the corresponding average changes in the accuracy percentages of the gait parameters measured between the start and end of the long walk. The overall drift across all gait parameters is 4.89\%, indicating measurement performance of our device remains accurate with low drift after prolonged walks. This is attributed to the stable device installation, low drift in camera parameters described in Section II, and the system's high resistance to environmental changes. Here we compared gait symmetry in terms of gait length and width, emphasizing relative changes rather than absolute values. \\
Some of the findings are consistent with the results from the first experiment, such as the larger absolute error in stride length due to the accumulation of errors. Besides, all spatial gait parameters exhibit higher absolute variation errors compared to temporal parameters. This is attributable to the substantially larger baseline values of spatial parameters relative to temporal parameters. Meanwhile, each temporal gait parameter exhibits low drift, with the exception of double support time, which shows a higher degree of drift. This increased drift is also observed in both the variation and symmetry measures. The underlying cause is that these parameters' calculations depend on other metrics, such as gait length, stride time, and swing/standing time, leading to the propagation of drift in subsequent calculations. A potential solution could involve adding a feedforward loop to account for the previous drift when calculating the subsequent dependent parameters.

\subsection{Person Identification Prediction Accuracy}
In Fig. 14(c), we present the data points collected by the designed gait monitoring system from the first three participants, focusing on gait length, stride time, and step velocity. These three parameters are commonly used in clinical settings to assess patients' conditions. The figure clearly shows three close clusters that can be easily differentiated. However, identifying such differences in gait is challenging for doctors or nurses using only visual observation or simple timing methods, as these participants appear to have the 'same gait pattern.' From the figure, it is evident that our hardware can detect subtle variations in gait, allowing for easy distinction between participants.

Ten parameters were utilized in the training process, including gait height, gait width, gait length, length symmetry, length variation, step time, stride time, swing time, double support time, and velocity. K-fold cross-validation was applied to evaluate the accuracy of the trained Transformer model. In each fold, 90\% of the data was allocated to the training set, while the remaining 10\% served as the validation set. Fig. 14(d) displays the confusion matrix of the predictions made by the trained Transformer. The model achieves an average accuracy of 95.7\%, indicating that each participant exhibits a distinguishable gait pattern. This suggests that our hardware may potentially identify patterns associated with gait disease, such as Parkinson's disease. The prediction for Participant 2 shows a higher error rate compared to the others. This error may be attributed to data in the validation set that was collected at the beginning or end of each trial. For certain participants, there is a noticeable difference in walking patterns between the starting/stopping phases and the middle, steady walking phase.
\begin{figure*}[!t]
\centering
\includegraphics[width=7.17in, height=1.59in]{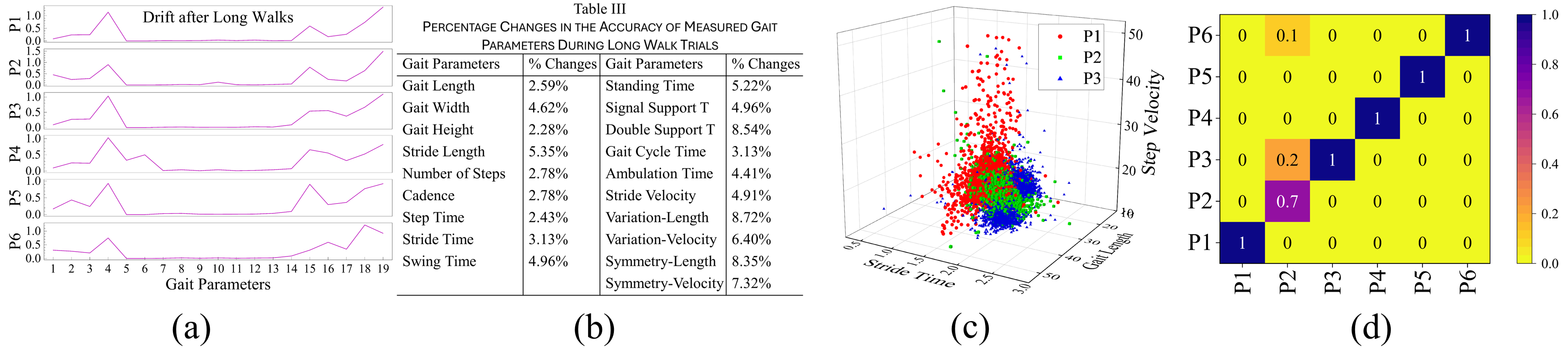}
\caption{(a) Average absolute variations across 19 gait parameters (x-axis label order follows that of Fig. 14(b)) between the start and end of the long walk; (b) Percentage changes in the accuracy of measured gait parameters during long walk trials; (c) 3D coordinate map (x: stride time, y: gait length, z: step velocity) displaying gait data points of the first three participants; (d) Confusion matrix showing the results of the person identification task (x: actual, y: prediction).}
\label{fig_14}
\end{figure*}

\section{Discussion}
Several points in this work are worth discussing. One point is that although the camera we use is compact, the small distance between the stereo camera lenses poses challenges for accurate triangulation. This issue could be addressed by redesigning the camera platform to increase the lens separation. Moreover, during the experiments, the devices were positioned parallel to the ground, and the angle between the camera's direction and the walking path was measured in advance. However, walking could cause the camera platform to shift, introducing errors in the measurement of gait parameters. This issue may be mitigated by integrating a gyroscope into the camera platform, enabling real-time detection and adjustment of the camera's position. Most importantly, the current gait monitoring system cannot measure certain parameters that the gait mat provides, such as double support loading, double support unloading, and foot flat time. A feasible solution would be to add another FSR at the tip of the shoe, as implemented in our previous work. In this study, we opted to remove one FSR at the tip as a trade-off to enhance the safety of the wearable device by minimizing dangling wires. This issue can be addressed by developing an embedded system with smart insoles, enabling the capture of a broader range of gait parameters. Finally, the ground truth measurement for gait height could be improved by using a parallel, highly accurate motion capture system.

Future work focuses on addressing the aforementioned challenges, while also exploring several other aspects. A low-power embedded system will be developed to measure additional gait parameters. Also, we plan to transition from a controlled laboratory environment to real-world conditions and recruit more participants with diverse gait patterns and conditions. A more extensive and comprehensive gait dataset will be collected, and a novel Transformer model will be developed to classify and predict gait abnormalities. Furthermore, a 3D trajectory of foot movement will be reconstructed using the collected gait data.

\section{Conclusion}
In conclusion, we have proposed and demonstrated a novel stereo camera-based, shoe-mounted wearable hardware system for gait monitoring using computer vision. The system is capable of measuring up to 17 gait parameters with high accuracy, all exceeding 93.61\%. The measurements are highly reproducible and exhibit a low drift of 4.89\% during extended continuous walks. Based on these properties, an individual gait identification task performed on the dataset collected by the device, trained using Transformer models, achieved an accuracy of 95.7\%. This indicates that our hardware is an effective tool for collecting long-distance gait sequences, making it suitable for integration into current LLMs. The device is low-cost, user-friendly, and capable of real-time measurement, while delivering competitive performance compared to state-of-the-art technologies.

\section{Acknowledgment}
The author would like to express gratitude to Dr. Jianwei Ke (Apple Inc.) for valuable discussions on the concept.

\vfill

\end{document}